\documentclass[a4paper,11pt]{article}

\pdfoutput=1 
\usepackage{jinstpub} 

\title{\boldmath Ce:LaBr$_3$ crystals with SiPM array readout and
temperature control for the FAMU experiment at RAL}

\author[a,1]{M.Bonesini,\note{Corresponding author.}}
\author[a]{R.Benocci,}
\author[a]{R.Bertoni,}
\author[a]{M.Clemenza,}
\author[a]{D.Ghittori,}
\author[a]{R.Mazza,}
\author[a]{E.Vallazza,}
\author[b]{A.deBari,}
\author[b]{A.Menegolli,}
\author[b]{M.Prata,}
\author[b]{M.Rossella,}
\author[c,2]{M.Baruzzo,\note{also at Dipartimento di Scienze Informatiche,
Matematiche e Fisiche, Universit\`a di Udine.}}
\author[c]{E.Mocchiutti}

\affiliation[a]{Sezione INFN Milano Bicocca, Dipartimento di Fisica 
G. Occhialini and Dipartimento di Scienze dell' Ambiente e della Terra, Universit\`a di Milano Bicocca, Milano, Italy}
\affiliation[b]{Sezione INFN Pavia and  Dipartimento di Fisica, Pavia, Italy}
\affiliation[c]{Sezione INFN Trieste, Trieste, Italy}

\emailAdd{maurizio.bonesini@mib.infn.it}

\abstract{ 
 Compact X-rays detectors made of  1/2'' Ce:LaBr$_3$ crystals of cubic shape
with SiPM array readout have been developed for the FAMU experiment at
RIKEN-RAL, to instrument regions of difficult access. Due to the high 
photon yield of Ce:LaBr$_3$  it was possible to use a simple readout
scheme based on CAEN V1730 digitizers, without a dedicated amplification
stage. 
The drift with temperature of 
SiPM gain was 
corrected by using  CAEN A7885D regulated 
power supply chips with temperature feedback.
Energy resolutions (FWHM) around  $3.5 \%$ at the 
$^{137}$Cs peak  and around $9 \%$ at the $^{57}$Co peak were obtained. }


\begin{document}
\maketitle
\flushbottom

\section{Introduction}
\label{sec:intro}
The FAMU (\underline{F}isica degli \underline{A}tomi \underline{Mu}onici)
 experiment at RAL \cite{famu} 
is designed to measure the hyperfine splitting (HFS) in the ground
state (1S) of the muonic hydrogen. It aims at  a high accuracy 
determination of the proton Zemach radius \cite{zemach}, \cite{bakalov}. 
This experiment may  contribute to solve the so-called ``proton 
radius puzzle'': a large and still unsolved 
  disagreement between the proton charge
as measured with electrons or muons \cite{pohl}. 

A high intensity pulsed low-energy muon beam, stopping in a hydrogen
target, is used to produce muonic hydrogen (in a mixture of singlet F=0 and triplet
F=1 states). 
A tunable mid-IR (MIR) pulsed high power laser
then  excites the hyperfine splitting (HFS) transition of the 1S muonic
hydrogen (from F=0 to F=1 states).
Making use of  the muon transfer from muonic hydrogen to another
higher-Z gas in the target (such as $O_2$),
the $(\mu^{-}p)_{1S}$ HFS transition will be recognized by
an increase of the number of
 X-rays from the $(\mu Z^{*})$ cascade, during a laser  
frequency scan around the resonance value  $\nu_{0}$ ($\Delta E_{HFS}=h \nu_{0}$) .
From the measurement of $\Delta E_{HFS}(\mu^{-}p)_{1S}$  the Zemach radius $r_Z$ of the proton
 may be computed  with a precision up to $5 \times 10^{-3}$,
thus  casting  new light on 
the proton radius puzzle.

\begin{figure}[htbp]
\centering 
\includegraphics[width=.54\textwidth]{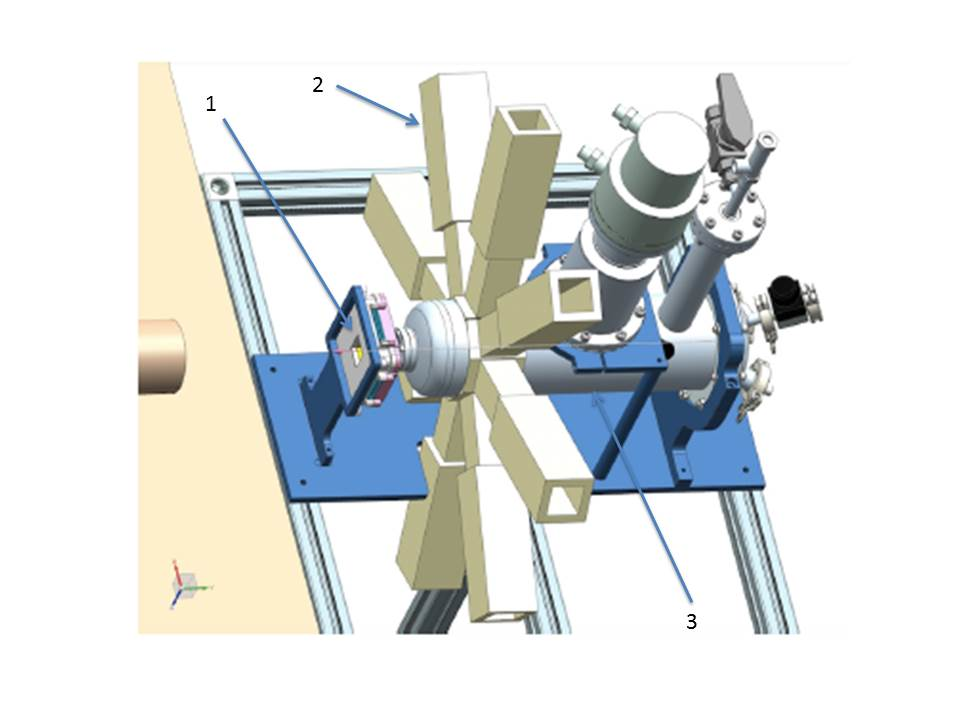}
\includegraphics[width=.44\textwidth]{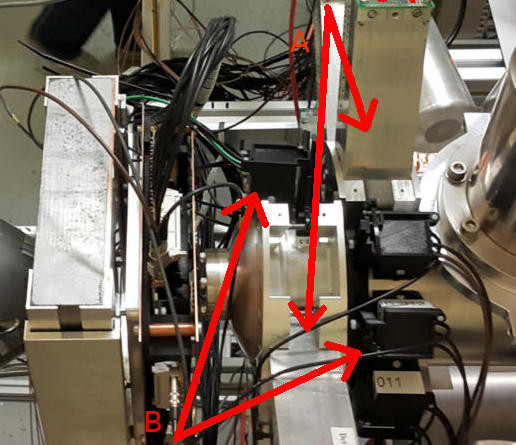}
\caption{Left panel:
 layout of the setup for the 2015-2016 data-taking (R582); 1)
is the 1 mm pitch beam hodoscope, 2) the crown of eight Ce:LaBr$_3$ detectors with PMT readout
and 3) the
cryogenic target. The four HPGe detectors, also used in this run, are not 
shown. 
Right panel: picture of the 2018 setup where the two half-crown of the Ce:LaBr$_3 $
detectors with PMT readout (A) were displaced along the beam axis (z) and 
complemented with 4+4  1/2"
Ce:LaBr$_3$ detectors with SiPM array readout (B).  }
\label{fig1}
\end{figure}

The FAMU experiment is   performed in steps, starting from the study of
the transfer rate from muonic hydrogen to another higher-Z gas
and ending with the  full working setup
including the pump MIR laser and a multipass optical cavity~\footnote{to 
enhance the probability of laser light-muon interactions}.
The preliminary steps have allowed to
determine the best mixture to be used inside the cryogenic target and
optimize the operating conditions.
A schematic layout of the experimental setup for the
preliminary steps is shown in figure \ref{fig1}.

The RIKEN-RAL muon facility \cite{riken} 
at Rutherford Appleton Laboratory (UK) provides
high intensity pulsed muon beams at four experimental ports. 
The  primary proton beam at 800 MeV/c 
impinges  on a secondary carbon target producing pions and then
high intensity
low energy pulsed muon beams. 
The muon beams reflect the primary beam structure: two pulses with a 70 ns 
FWHM and a 320 ns peak to peak distance are delivered, with a 50 Hz repetition 
rate. The FAMU experiment
makes use of a negative 
decay muon beam at $\sim$ 60 MeV/c. 
For this experiment, an important issue is the optimal steering of the
incoming high intensity pulsed muon beam onto the hydrogen target, to
maximize the muonic hydrogen production rate.
A system of three beam hodoscopes has been developed for this scope. 
The first two are based on square $3 \times 3$ mm$^2$ Bicron BCF12 
scintillating fibers read 
by SiPMs, while
the last one is based on square $1 \times 1$ mm$^2$ scintillating fibers
of the same type, with white EMA coating, to avoid light cross-talk
 \cite{carbone}. 
The muon beam intensity is around 
$6 \times 10^{4} \ \mu^{-}$/s in a typical size $4 \times 4$ cm$^2$. 
The energy spread is around $10 \ \%$ and the angular divergence around 
60 mrad. 

To extract the characteristic 
muonic X-rays lines (around 100 keV) with a good energy 
resolution and a minimal events pile-up, a system based on Ce:LaBr$_3$ crystals
and HPGe detectors has been developed. Even if they have better energy 
resolution, the HPGe detectors are slower, work at cryogenic temperatures and
are more expensive. Therefore the main X-rays detector system for the experiment
was based on 1" circular Ce:LaBr$_3$ crystals, 1'' long,  read by UBA Hamamatsu R11265U-200
 PMTs with active divider (up to eight arranged
in two  detachable half crowns). 
In addition, an R $\&$ D was pursued to complement these detectors with crystals
equipped with SiPM readout to instrument regions of more difficult access,
see reference \cite{adamczack18} for further details.
 
\section{ X-rays detectors with SiPM arrays readout}
\label{sec:constr}
For our aims it is essential to detect low-energy X-rays
in the range 100-200 keV. Pr:LuAG~\cite{prluag} and Ce:GAAG~\cite{cecaag}
crystals with  respect to more conventional
Ce:LaBr$_3$, CeBr$_3$\cite{cebr3} and NaI(Tl) crystals, have the  advantage  to be non
hygroscopic and thus do not need
encapsulation.

Results on their performances are reported in references \cite{bonesini15}, 
\cite{bonesini17}. 
From laboratory tests a solution based on Ce:LaBr$_3$ crystals 
was shown as still to be preferred.
A crystal thickness of 0.33 (1.54) cm for $88 \%$ attenuation  at 100 (200)
keV was  computed from X-ray attenuation coefficients,
as reported in \cite{nist}. It is apparent that for the detection of the
O$_2$ characteristic lines in the region 100-160 keV, corresponding to muon 
transfer, 1/2'' long
crystals are adequate. 
A more complete Monte Carlo simulation based on MNCP\cite{mncp}
  provided an estimate of absorption for
cubic crystals of 1/2'' side with a source at a distance corresponding to the
center of the foreseen target. Even in this case 1/2'' long crystals were
considered 
adequate.  

\begin{figure}[htbp]
\centering 
\includegraphics[width=.85\textwidth]{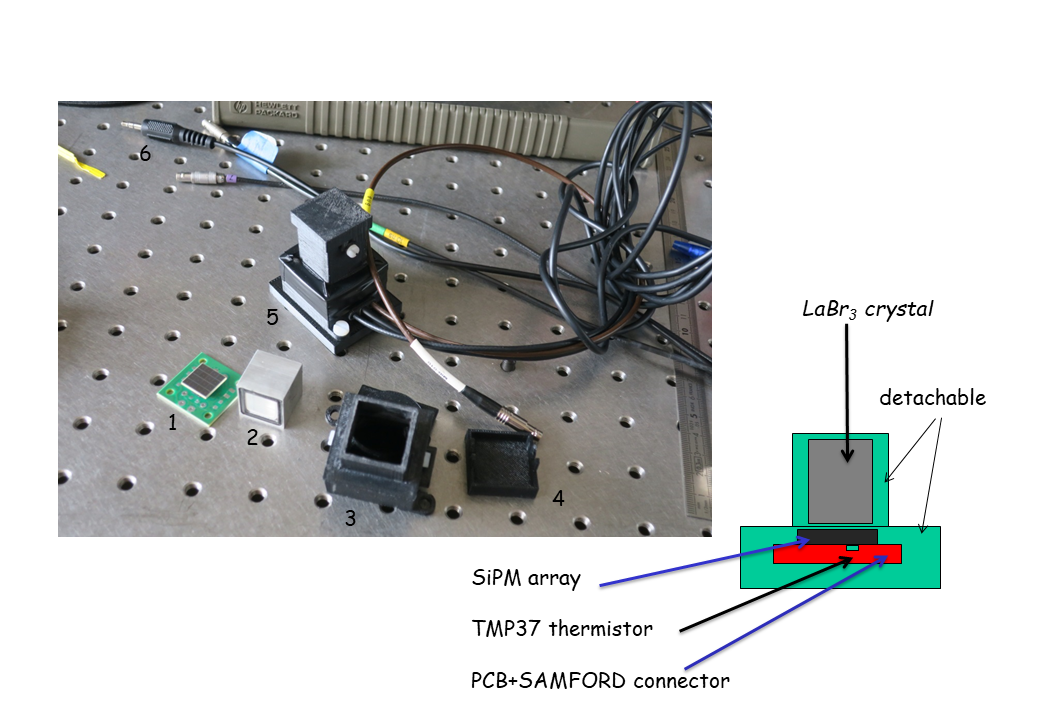}
\caption{Components of the detectors with SiPM readout. 1) is the Hamamatsu SiPM mounted on the custom PCB, 2) the Ce:LaBr$_3$ crystal in the Al 
encapsulation, the optical window is seen in the front. 
3) is the holder containing the crystals and the PCB, in two
pieces seen from the top. 4) is the cap to guarantee detector's light tightness. 
The full mounted detector (5) is shown
in the back of the picture. The 3.5 mm stereo jack cable (6) connects the 
temperature sensor (TMP37) on the backside of the SiPM array PCB to the 
power supply module.}  
\label{fig2}
\end{figure}

The structure of a detector with a SiPM array readout is shown in figure 
\ref{fig2}. The optical contact between the crystal and the SiPM arrays is
done through a Bicron BC631 silicone optical grease. 
The crystal/PCB holder, realized with a 
3D printer,  is made  in two
pieces: one contains the crystal under test, while the other holds the 
readout electronics with the custom PCB, where a SiPM array is mounted on. 
The analog signals of the 16 SiPM of one array are summed together on the
custom PCB. 
Signal acquisition may be realized with a standard spectroscopic chain 
(based on a Ortec 672 spectroscopic amplifier or a fast Ortec 579 Ortec
amplifier) or with a fast digitizer. Due to the signal amplitude ($\sim 100-
200$ mV at the $^{137}$Cs peak ) no amplification is needed and a direct
readout via a digitizer may be used. In our case we made use of a CAEN
DT5730 digitizer (desktop version) or a V1730 digitizer (VME format). 
Both have been used via an optical link and have a bandwith of 500
MHz with a $\pm 1$ Vpp dynamic range. 

In the first instance, we  used different 
 4 $\times$ 4 
arrays made from 3 $\times$ 3 mm$^2$ SiPM array from Sensl, Advansid and Hamamatsu 
for their readout. Their main  operation characteristics  are resumed  
in table \ref{tab2}. 
Hamamatsu SiPM make use of the TSV (``through Silicon via'') technology that
eliminates the need of a wire bonding pad, thus reducing  dead space 
problems. The anode of each channel is traced to the backside pad by TSV. 
Typical gains are in the range 1.7 to 3 $\times 10^6$ and depend on the 
applied overvoltages ($\Delta V_{ov}$), 
while the dark count rate is around 0.5 Mcps for all the considered SiPM arrays. 
\begin{table}[htbp]
\caption{Main characteristics of the SiPM arrays used for our tests.Photon detection efficiency 
(PDE) are at typical overvoltage values, at $\lambda_{max}$. 
$V_{op}=V_{bd}+\Delta V_{ov}$ is the typical voltage used in our 
tests. $\Delta V_{op}$ is the variation in the suggested voltages for operations, between the
	16 different SiPM making a SiPM array. (E) or (S) are used for
an epoxy or silicon window.
}
\label{tab2}
\smallskip
\centering
\begin{tabular}{|l|c|c|c|c|c|c|c|c|}
\hline
             & $V_{bd}$ & $\Delta V_{ov}$ & $V_{op} $ & $\Delta V_{{bd}}/\Delta T$  & $\Delta V_{op}$ & $\lambda_{max} $ & PDE  & range    \\
             &  (V)     &   (V)      &  (V)             & (mV/C) & (V)  &  (nm)  &($\sim \lambda_{max})$   & (nm)  \\\hline
SenSL Array   &  24-25& 1-5   &26  & 21.5 &   &   420 & $\sim 30 \%$ & 300-800 \\
SB-4-3035-CER&      & & &    &       &           &         &             \\ \hline 
Advansid     & 26   & 2-6    & 29  & 26 & $\leq 0.4$& 420 & $\sim 43 \%$ & 350-900  \\
NUV3S-4x4TD      &  &     &            &    &     & &              &          \\ \hline 
Hamamatsu   & $53 \pm 5$ & $\sim 3$  & 53.8 & 54 & $\pm 0.05$ & 450 & $\sim 35 \%$ & 320-900 \\
	S13361-3050-AE (E)   &  &   &      &    &     &              &     &        \\ \hline
Hamamatsu  & $53 \pm 5$ & $\sim 3$  & 54.2 &54 & $\pm 0.05 $ & 450 & $\sim 35 \%$ & 280-900 \\
	S13361-3050-AS (S)  & & &      &   &     &              &        &    \\\hline
Hamamatsu & $38$ & $\sim 2.7$ & 40.8 & 34 & $\pm 0.05$ & 450 & $\sim 50 \% $
	& 270-900 \\
	S14161-3050-HS (S) & & & & & & & & \\	
\hline
\end{tabular}
\end{table}

Preliminary results obtained with a standard spectroscopic chain 
were reported in references \cite{bonesini15}, \cite{bonesini17}  and 
show resolution 
 at 662 keV  
from $3.1 \%$ (Ce:LaBr$_3$ crystals with Hamamatsu S13361-3050-AS SiPM arrays) to
8.4 $\%$ (NaI crystals with the same readout). 
At lower X-rays energy ($\sim 122$ keV), FWHM energy resolutions between the different
crystals become more compatible: as an example while at 662 keV a Ce:LaBr$_3$ 
crystal has  a resolution a factor $\sim 2$ better than a NaI(Tl) crystal, 
at 122 keV this 
factor reduces only to $\sim 30 \%$ .

The best results were obtained with Ce:LaBr$_3$ crystals with a readout based
on Hamamatsu  SiPM arrays with a silicone window,
that has a better transmission around 380 nm.

As the SiPM gain has a drift with temperature ($\sim 54 $ mV/$^{\circ}$C 
for the 
breakdown voltage of Hamamatsu S13361 SiPM) a temperature   correction
had to be implemented.

\subsection{Temperature control of SiPM gain}
\label{sec:temp}
The gain of SiPM depends on the applied voltages $V_{op}=V_{bd}+ \Delta V_{ov}$
where the overvoltage is kept fixed: typically around 2-4 V. The breakdown
voltage depends from temperature,  
 according to equation: 
 \begin{equation}
 \label{eq1}
V_{bd}(T)= V_{bd}(T_0)+c \times (T-T_0) 
 \end{equation}
where $c$ is the temperature coefficient $\Delta V_{bd}/\Delta T$ of table
\ref{tab2} and $T_0$ a reference temperature, tipically 25 $^{\circ}C$ \cite{otte}. 
By correcting for the increase of breakdown 
voltage with the previous equation  \ref{eq1}
  one may obtain an excellent gain stabilization.  The temperature correction
  may be obtained offline, by recording the temperature, or online with an
  active feedback. 
\begin{figure}[htbp]
\centering 
\includegraphics[width=0.50\textwidth]{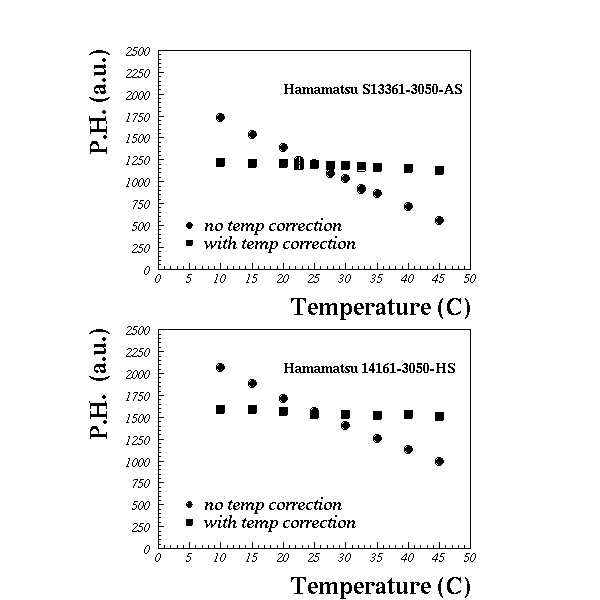}
\includegraphics[width=.49\textwidth]{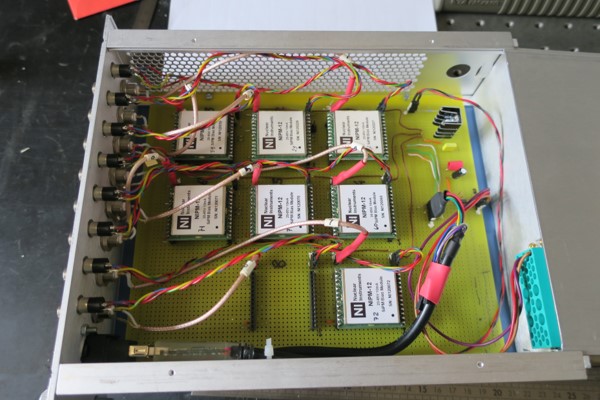}
\caption{Left panel: P.H. response in a.u. for a typical Ce:LaBr$_3$ 1/2'' crystal as a 
function of temperature with a $^{137}$Cs source. 
 Data have been taken inside an IPV30 Memmert climatic
	chamber, with a temperature resolution $\sim 0.1^{\circ}$C. Data have been taken 
with no temperature correction (circle) and with temperature correction 
(squares).
Right panel: top view of the custom NIM module for temperature
control of SiPM gain. The USB-I2C interface is shown in the bottom
part of the picture. Seven out of eight CAEN A7585D power supply
chips   are shown  in place.
A wire-wrap mounting has been used. }
\label{fig3}
\end{figure}

The effect to be corrected is shown in figure \ref{fig3} (black circles)
for two different Hamamatsu SiPM arrays. 
Data have been taken inside a Memmert IPV30 climatic chamber, with a
temperature resolution of 0.1 $^{\circ}$C between 10 $^{\circ}$C and 40 
$^{\circ}$C. 
A typical detector is irradiated with a $^{137}$Cs source and data are 
read by a CAEN V1730 digitizer. The position of the pulse height peak
is computed and then plotted as a function of temperature. 
We initially used single desktop power supply CAEN 
DT5485P~\footnote{ 0.1 mV (pp) voltage ripple, $\pm 20$ mV setting precision, 
1.2 mV setting resolution, with USB control}, 
where the temperature feedback was based   
on a temperature sensor (TMP37 from Analog Devices) put on the
backside of the PCB holding the SiPM
array (see figure \ref{fig2} for details). 
This temperature sensor is  connected via a 3.5 mm stereo cable 
to the power supply module.
Between 10 $^{\circ}$C and 40 $^{\circ}$C the detector pulse height response
had a variation up to 60 $\%$. This effect is reduced to
$\sim 6 \%$ after temperature correction, via a CAEN DT5485P desktop
module.

All  results, based on laboratory tests, were made using such 
modules for powering the SiPM arrays. For the next future we have 
developed 
 custom made NIM modules with up to eight HV channels each, based on CAEN 
A7585D  chips. The communication with the host computer is done via an 
I2C protocol, followed by an USB-I2C converter,
 using on the computer side a proprietary software \footnote{
Zeus software from Nuclear Instruments srl.} that may control 
the setting of the power supply
 chips, monitor their erogated voltages and currents 
and record results on an Excel file. These modules realized with a wire 
wrap technique are shown in figure \ref{fig3} and their use is foreseen
for the next spectroscopic run of FAMU in late 2020. 
In these modules the primary voltage  to feed the power supply
chips is taken 
from the NIM backplane and
the interface USB-I2C is realized via a FDTI C232HM-EDSHL-0 module.
For data taken
in December 2018 six  DT5485P CAEN modules connected  to an USB hub were 
used instead.
\section{Results from laboratory tests.}
\begin{figure}[htbp]
\centering 
\includegraphics[width=.48\textwidth]{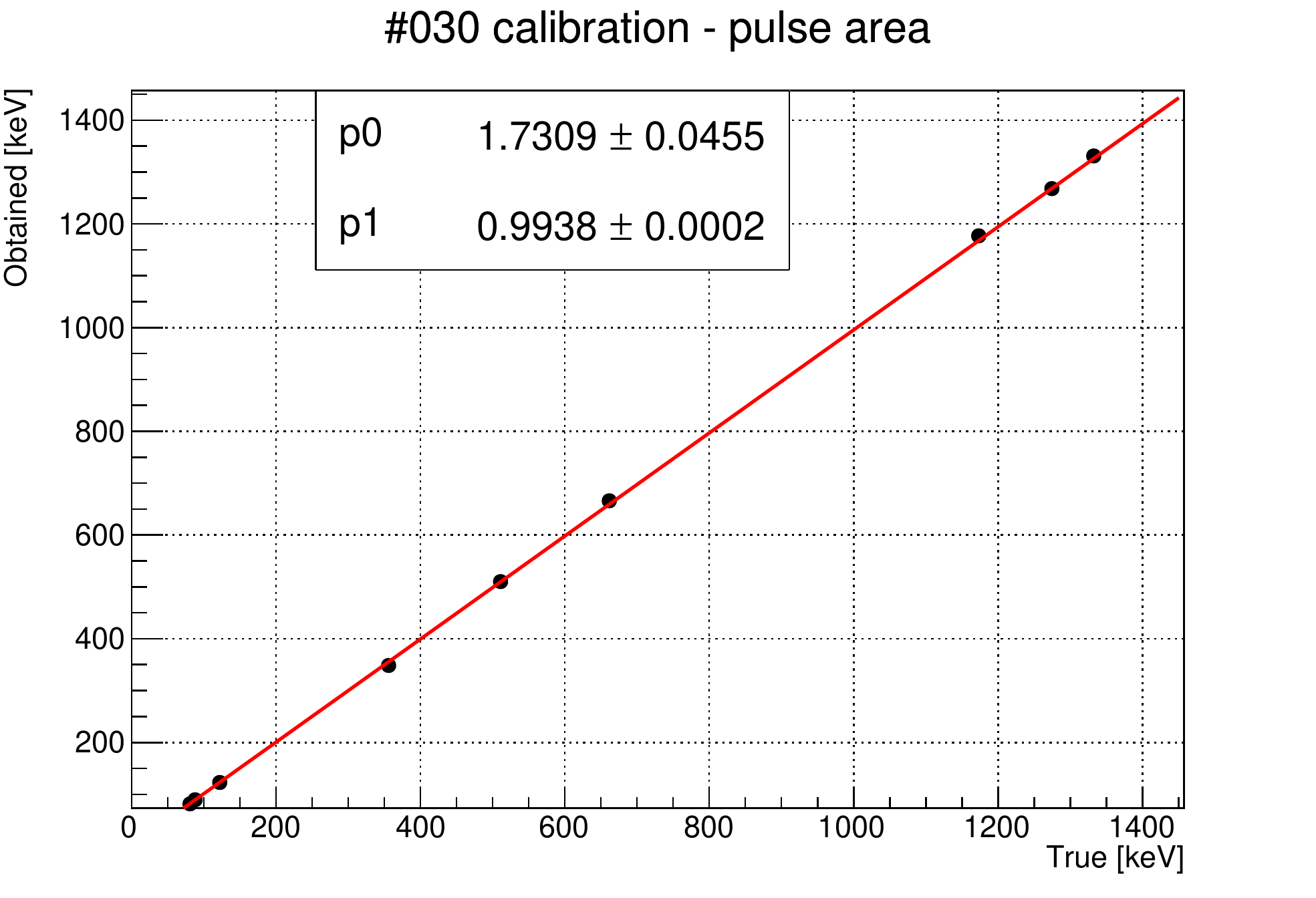}
\includegraphics[width=.48\textwidth]{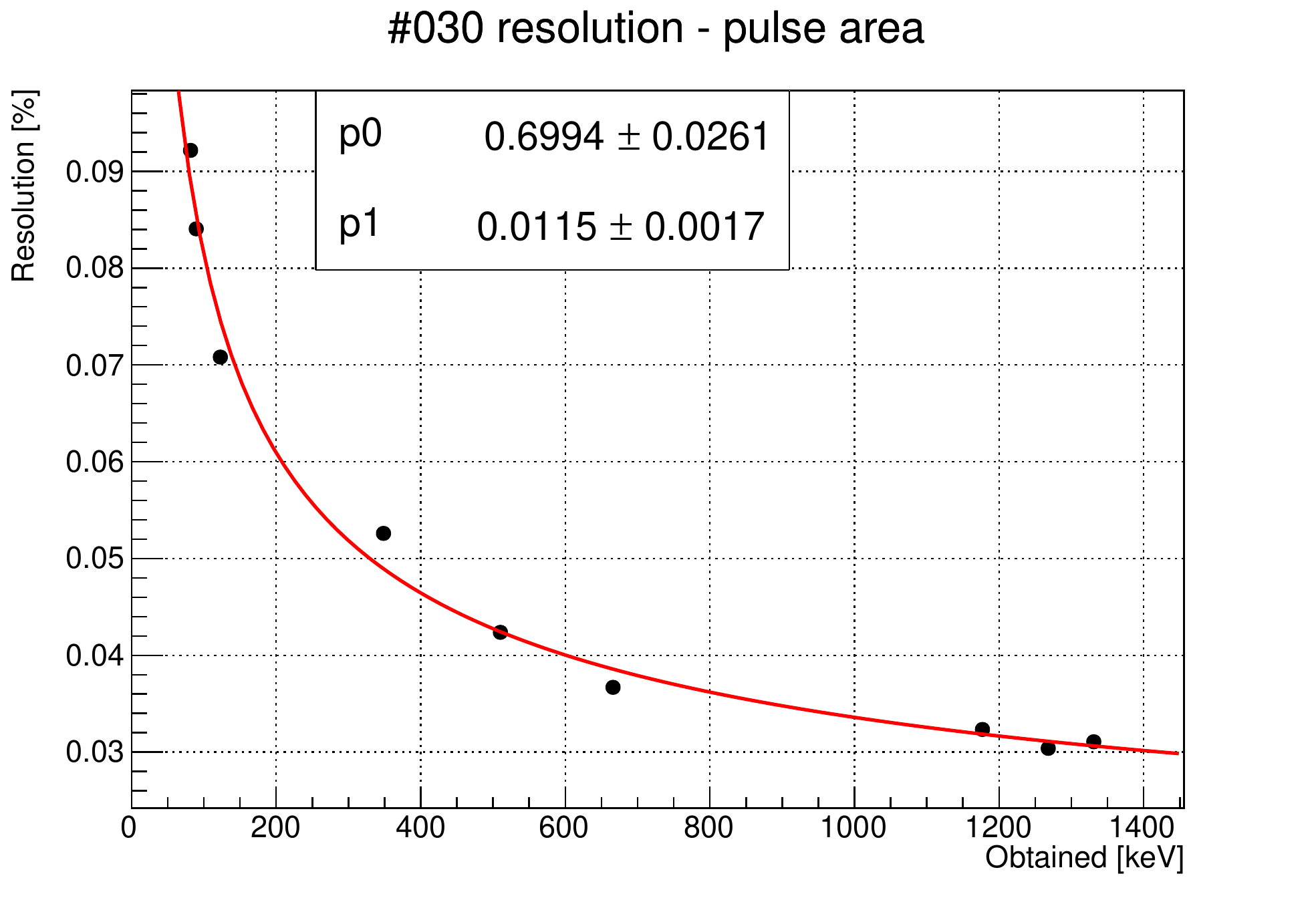}
\caption{Left panel: linearity for a typical 1/2" Ce:LaBr$_3$ detectors with
SiPM array readout, from 
OST Photonics (CN). Right panel:
FWHM resolution with different test sources for the same detector. Fits are 
performed with a straight line for linearity and with  the expression
$ p_0 + p_1/\sqrt(E)$ for the
FWHM energy resolution.}
\label{fig2c}
\end{figure}

\begin{figure}[htbp]
\centering 
\vskip -2cm 
\includegraphics[width=.5\textwidth]{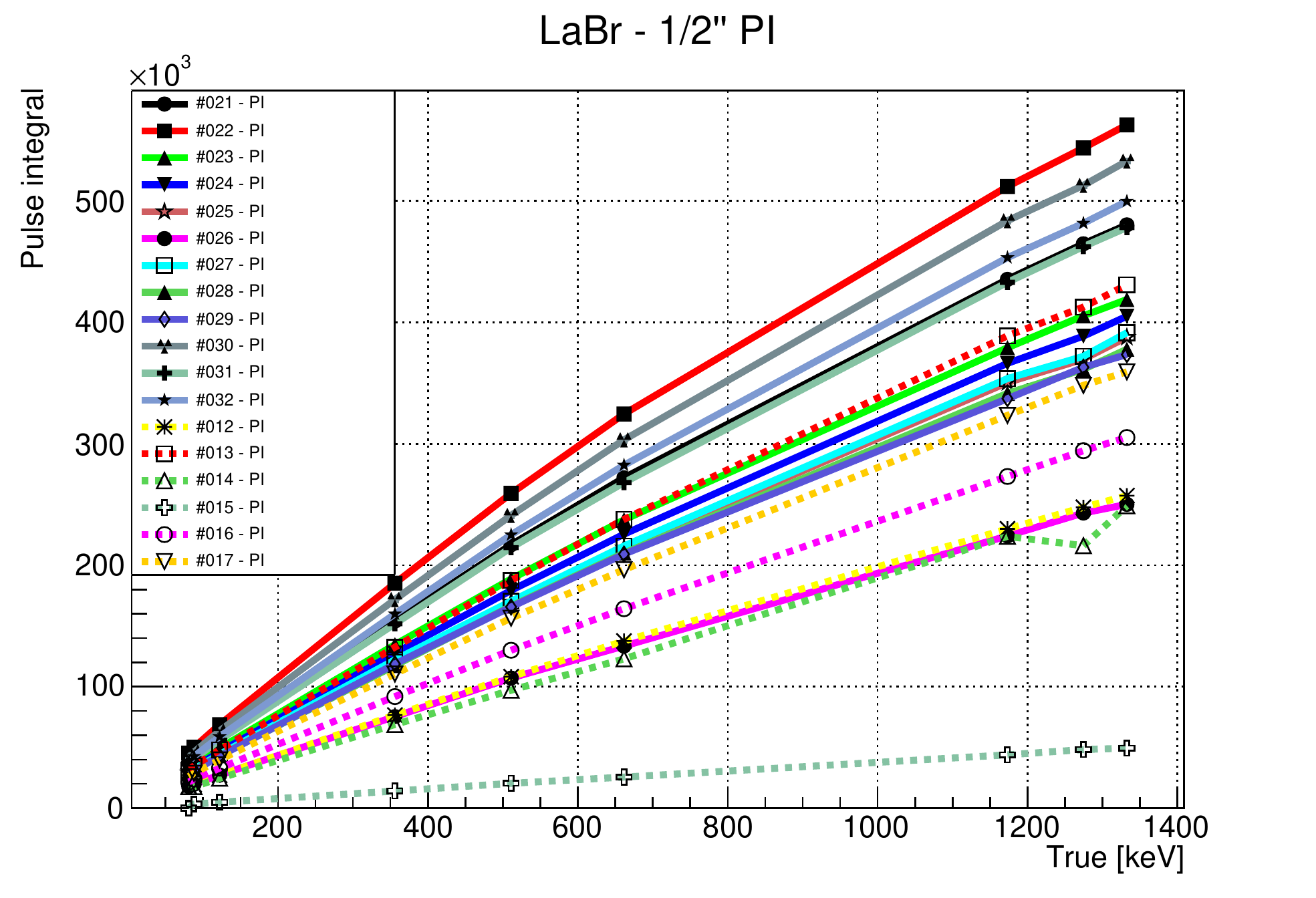}
\includegraphics[width=.5\textwidth]{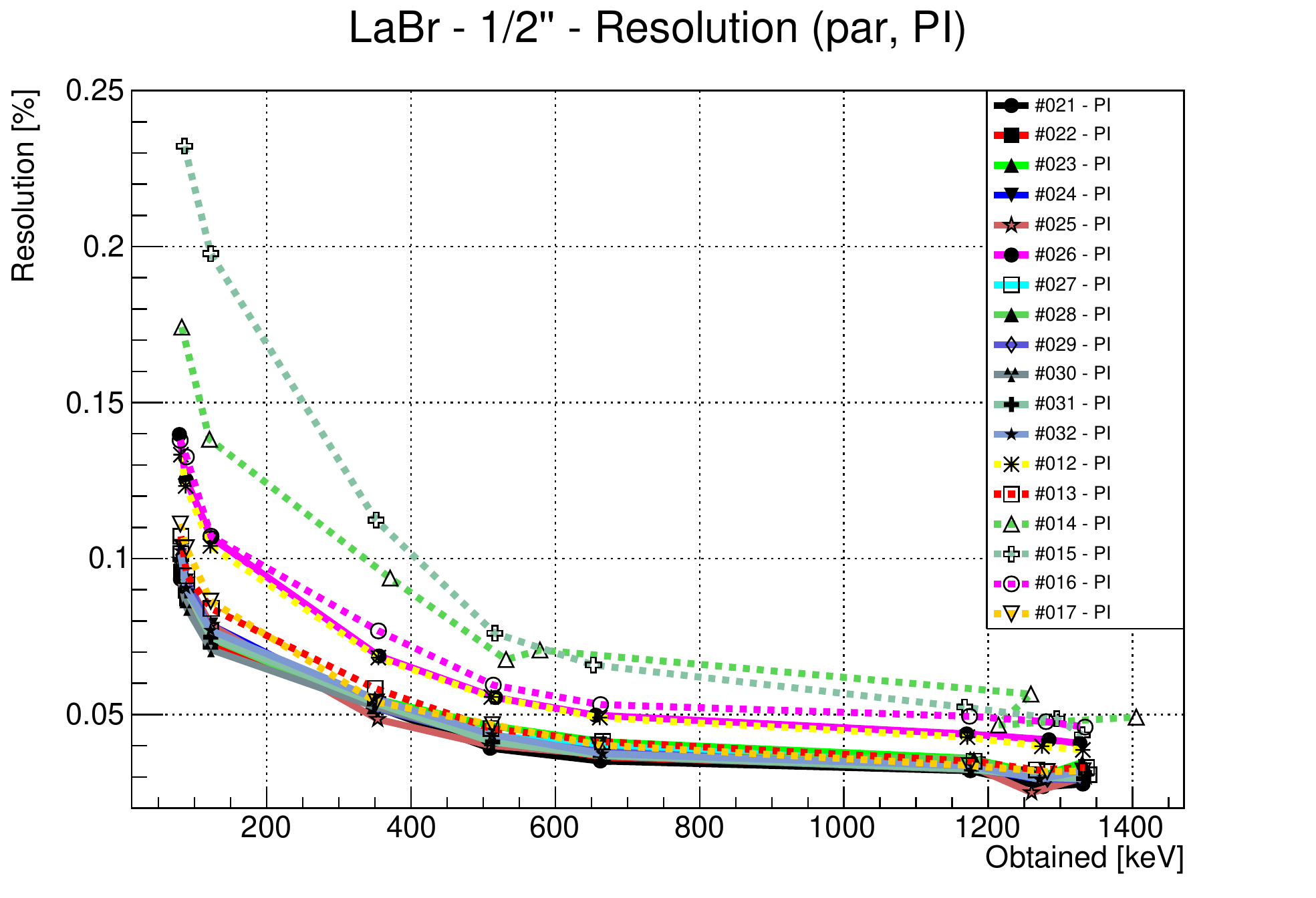}
\caption{Top panel: linearity for the used 1/2" Ce:LaBr$_3$ detectors with
SiPM array readout. Detectors no. 12-17 are from Kinheng Photonics (PRC),
while detectors no. 21-32 are from OST Photonics (PRC). Bottom panel:
FWHM resolutions with different test sources for the same detectors.}
\label{fig2b}
\end{figure}

Results for a typical crystal are shown in figure \ref{fig2c} for both
linearity and FWHM resolution (in $\%$) using different laboratory sources
in the range between 80 and 1300 keV. At the $^{137}$Cs peak a resolution
$\sim 3.5 \%$ was found, compatible with best results with the  standard 
PMT readout. 
Results on linearity and FWHM resolution ($\%$) are also shown 
in figure \ref{fig2b}
for several Ce:LaBr$_3$ detectors with size $14 \times 14 \times 14$ mm$^3$ from
Kinheng Ltd (no. 12-17) and $12 \times 12 \times 12$ mm$^3$ from Ost Photonics
(no. 21-32). 
Detectors with worse energy resolutions are equipped with Hamamatsu 
S13361 arrays with epoxy windows, that have a reduced transparency to
 the UV  signal emitted from Ce:LaBr$_3$ or have suffered from ageing problems (such as
some older detectors from Kinheng, PRC).
Around 122 keV FWHM energy resolutions up to $8 \%$ are obtained.  

\section{Analysis of performances in beam}
\label{sec:perform}
In the December 2018 run at Port 1 of RIKEN RAL, the two half-crown of 1''
Ce:LaBr$_3$ crystals with PMT readout were displaced of $\sim 10$ cm along 
the beam direction. They were complemented with four 1/2" Ce:LaBr$_3$ detectors
with SiPM readout, each. 
The first six detectors  were powered
via CAEN DT5475 modules with temperature feedback, the last two  were powered by conventional
ISEG NIM NHS-6001x power supply~\footnote{ with a voltage ripple less than
2-3 mV, 
 0.2 mV resolution voltage setting}  for cross-check. As the temperature in the experimental hall
was quite stable (the run was done in winter) no appreciable temperature 
excursions were seen.
 
\begin{figure}[htbp]
\centering 
\includegraphics[width=.49\textwidth]{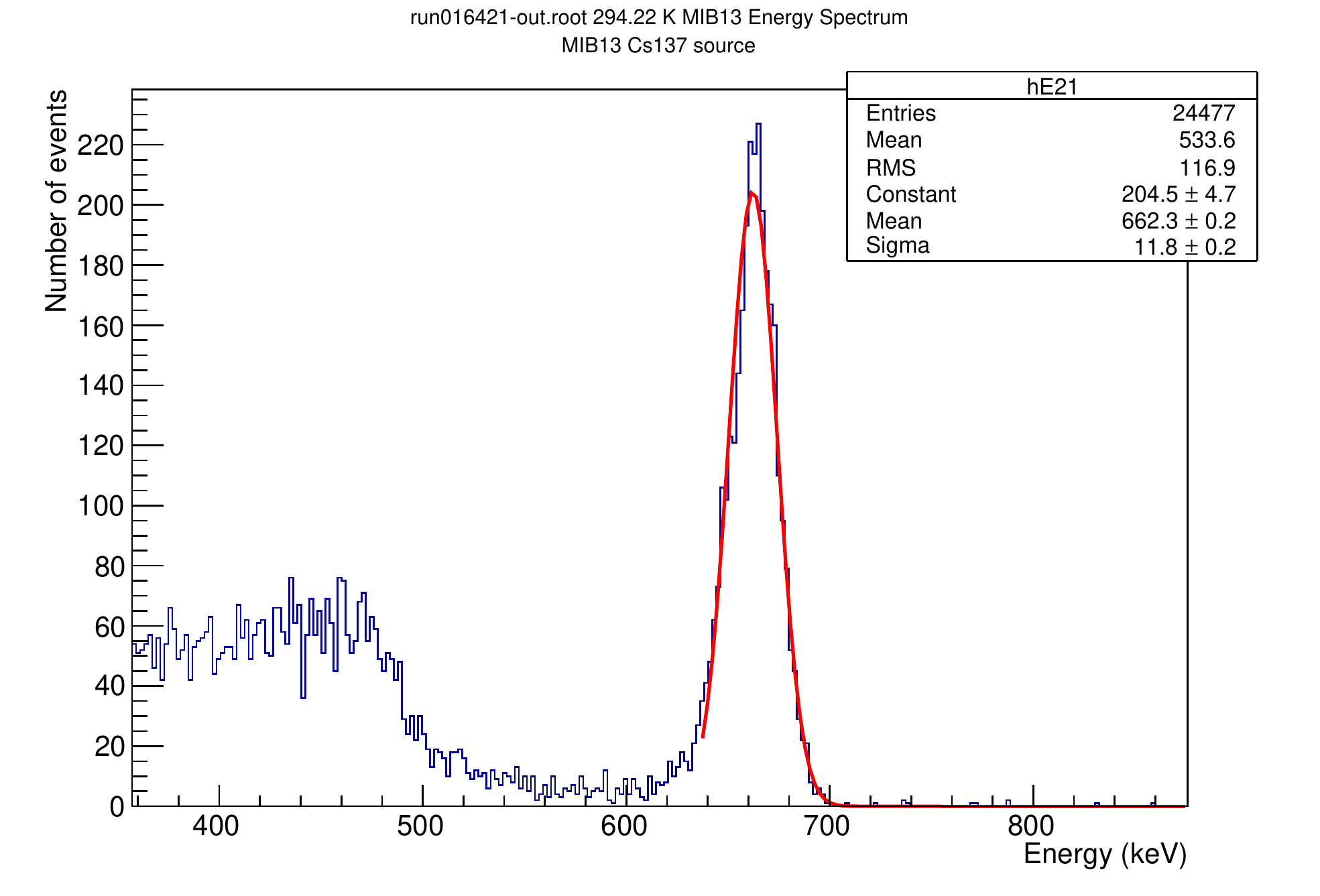}
\includegraphics[width=.49\textwidth]{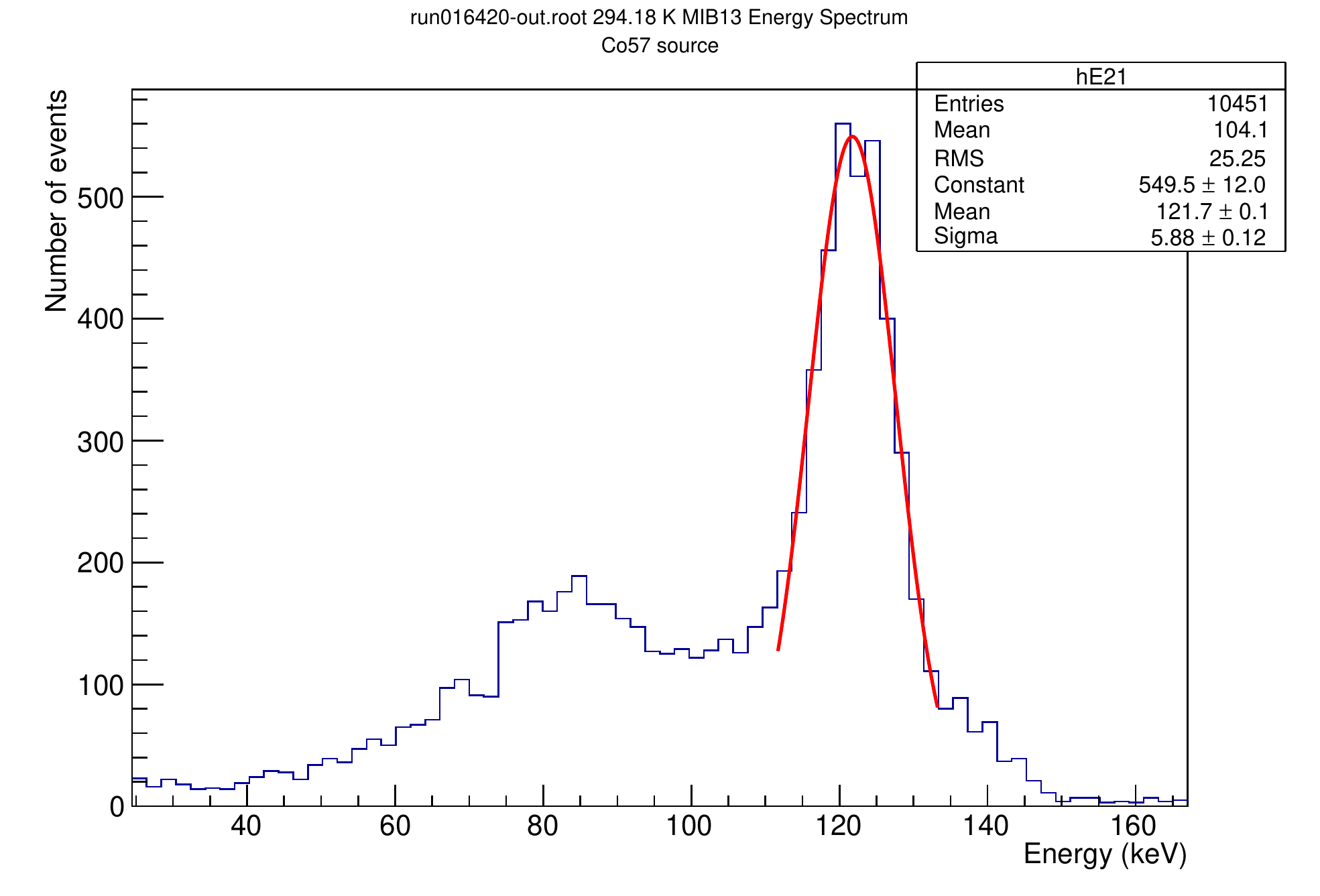}
\includegraphics[width=.49\textwidth]{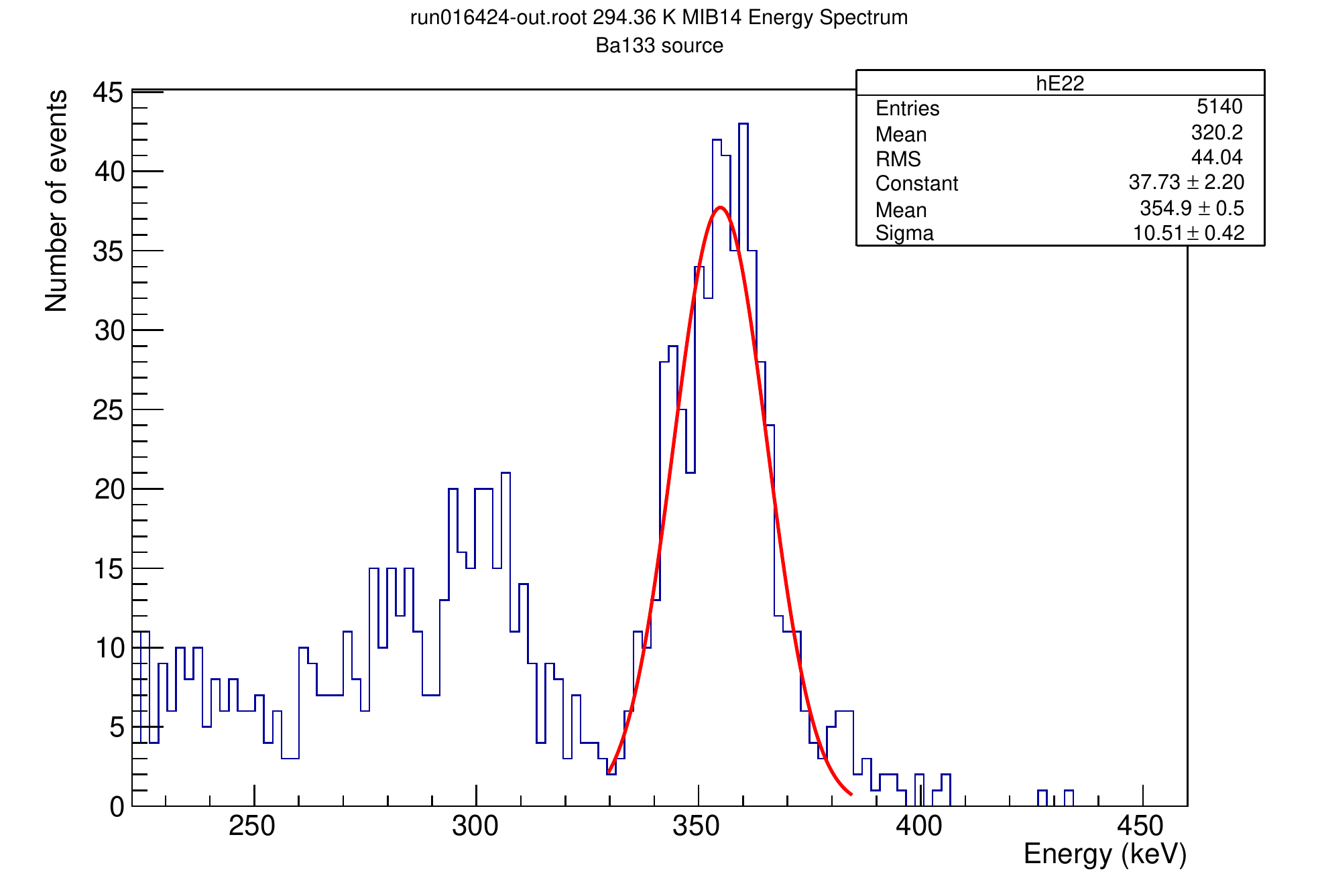}
\caption{Calibration results obtained with a $^{137}$Cs, $^{133}$Ba and
a $^{57}$Co source during the December 2018 run at RAL. FWHM resolutions 
are slightly worse as respect to  what obtained in laboratory measurements.}
\label{fig5}
\end{figure}

Calibration results in situ 
with $^{137}$Cs, $^{133}$Ba, $^{57}$Co sources are reported
in figure  \ref{fig5} and are roughly compatible with laboratory measurements,
even if FWHM resolutions are a little worse. This may be due to worse 
positioning of sources with respect to the detectors to be calibrated 
and environmental noise.

Data were then taken with a target filled with pure H$_2$ for background 
studies and a mixture of O$_2$ and H$_2$ at various concentrations 
(from 0.3 to 4.6 $\%$ weight) at a temperature 
around 80 K, at various pressures. 

\begin{figure}[htbp]
\centering 
\includegraphics[width=.75\textwidth]{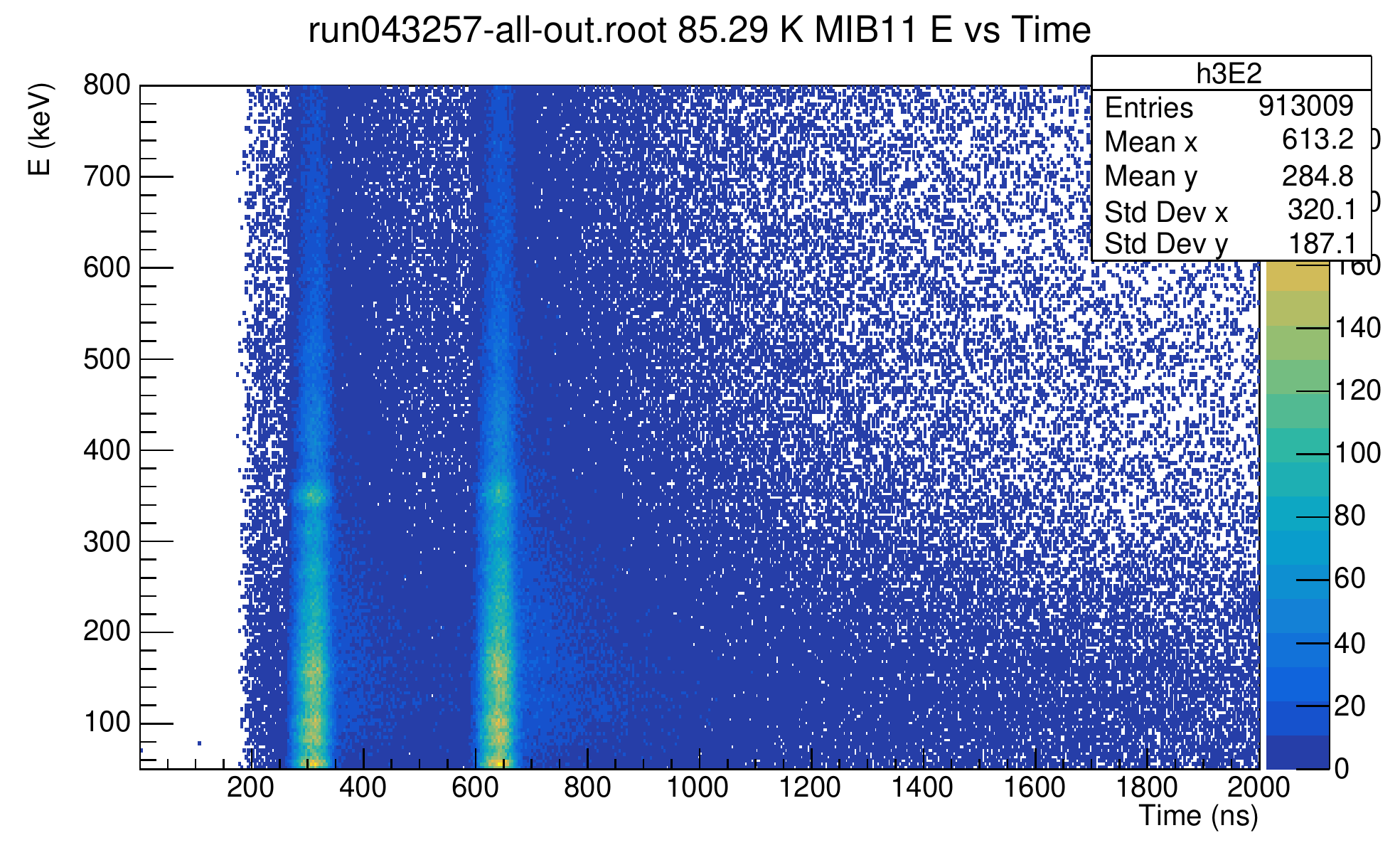}
\includegraphics[width=.75\textwidth]{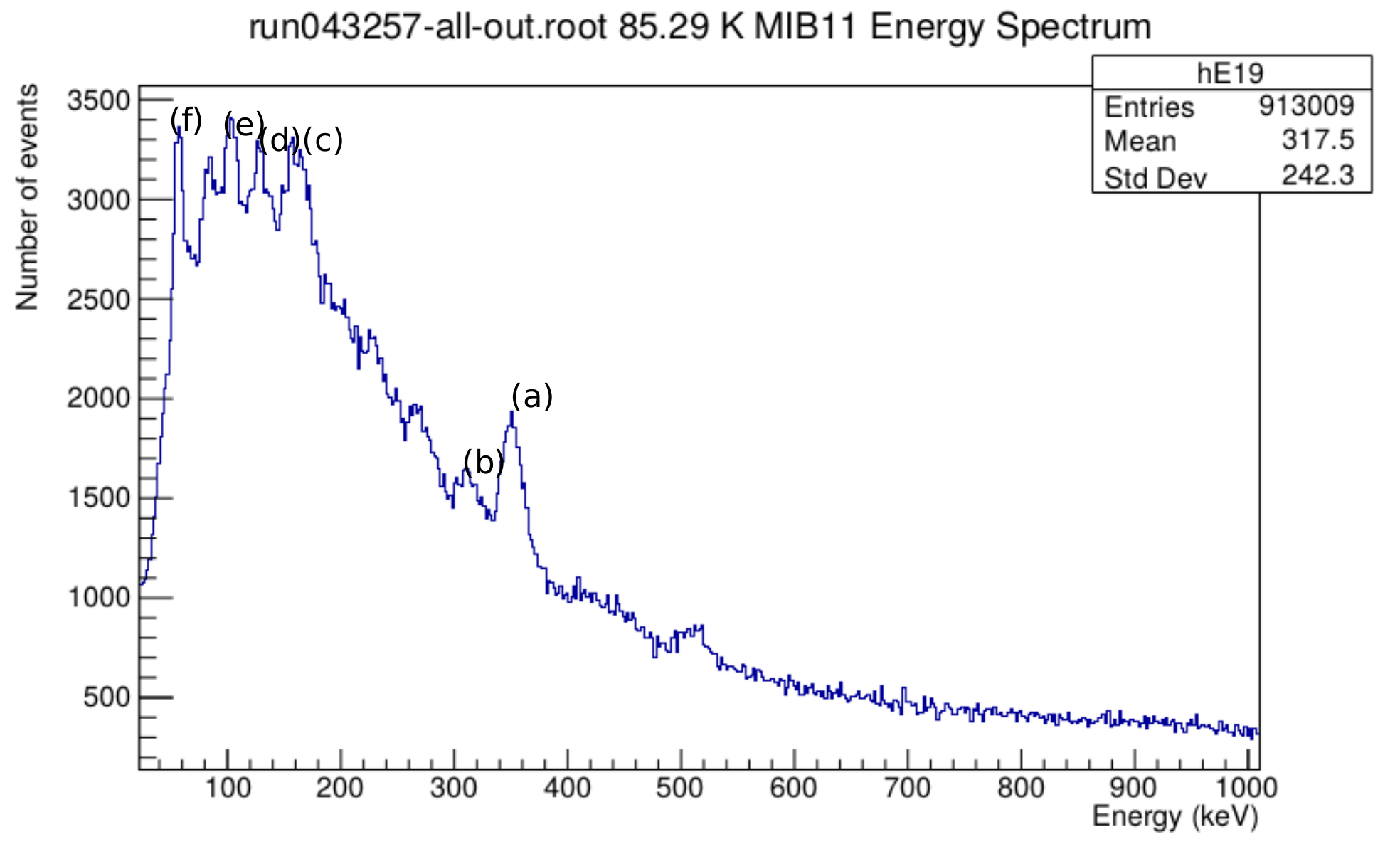}
\caption{ Top panel: X-ray time evolution spectrum of the H$_2$-2\% O$_2$ mixture with a 
56 MeV/c moun momentum.
Bottom panel:  energy spectrum for a SiPM array readout 1/2" Ce:LaBr$_3$ detector
using a 2 $\%$ $O_2$ mixture, at 7 bar and 80 K. Background is not subtracted. 
Characteristic spectral lines at 347 keV and 66 keV from Al (a,f), Nickel at
310 keV and 107 keV (b,e) and Oxygen at 158/167 keV 
and 133 keV (c,d) may be seen.}
\label{fig6}
\end{figure}

The timing properties of one typical detector  are shown 
in figure \ref{fig6}. The two peaks structure of the beam is clearly visible
with FWHM and peak-to-peak distance compatible with what expected.
 In figure \ref{fig6} the full energy spectrum is reported for the same 
detector. Characteristic X-rays lines, mainly Nickel and Aluminium from
materials present in the target,  are evident from $\sim 100$ keV to
around 400 keV. 

\section{Conclusions}
\label{sec:conc}
Ce:LaBr$_3$ detectors with SiPM array readout and temperature control 
for the power supply have been assembled and tested both in laboratory, 
at Sezione INFN Milano Bicocca and in beam at Port 1 at RIKEN RAL. 
Results show good perfomances and FWHM resolutions compatible with
more bulky conventional detectors with PMT readout. 
Their use is foreseen, together with the NIM power supply module
with temperature feedback, for the coming 2020 FAMU spectroscopic 
run. 

\section*{Acknowledgements}
We would like to thank S. Banfi, M. Gheigher (INFN Milano Bicocca)
for help in mechanics setup. We acknowledge
the help of Dr. M. Saviozzi of CAEN, Dr. A. Abba and Dr. V. Arosio of
Nuclear Instruments for issues related to the regulated power supply
chips for SiPM and their control program Zeus. 

\end{document}